\documentclass[prd,aps,twocolumn,showpacs,preprintnumbers,amsmath,amssymb,nofootinbib]{revtex4}
\usepackage[dvips]{graphicx}
\usepackage{epsf}
\usepackage{amsmath}
\usepackage{amssymb}

\usepackage{graphicx}
\usepackage{dcolumn}
\usepackage{bm}
\def\be{\begin{equation}}
\def\ee{\end{equation}}
\def\bea{\begin{eqnarray}}
\def\eea{\end{eqnarray}}

\usepackage{color}

\begin{document}


\title{Preheating a bouncing universe}

\author{Yi-Fu Cai$^{a}$, Robert Brandenberger$^{b}$ and Xinmin Zhang$^{c,d}$}

\affiliation{a Department of Physics, Arizona State University,
Tempe, AZ 85287, USA}

\affiliation{b Department of Physics, McGill University,
Montr\'eal, QC, H3A 2T8, Canada}

\affiliation{c Institute of High Energy Physics, Chinese Academy
of Sciences, P.O. Box 918-4, Beijing 100049, P.R. China}

\affiliation{d Theoretical Physics Center for Science Facilities
(TPCSF), Chinese Academy of Sciences, P.R. China}


\begin{abstract}
Preheating describes the stage of rapidly depositing the energy of cosmological scalar field into excitations of other light fields. This stage is characterized by exponential particle production due to the parametric resonance. We study this process in the frame of matter bounce cosmology. Our results show that the preheating process in bouncing cosmology is even more efficient than that in inflationary cosmology. In the limit of weak coupling, the period of preheating is doubled. For the case of normal coupling, the back-reaction of light fields can lead to thermalization before the bouncing point. The scenario of matter bounce curvaton could be tightly constrained due to a large coupling coefficient if the curvaton field is expected to preheat the universe directly. However, this concern can be greatly relaxed through the process of geometric preheating.
\end{abstract}

\maketitle

\newcommand{\eq}[2]{\begin{equation}\label{#1}{#2}\end{equation}}

\section{Introduction}

Non-singular bouncing cosmologies can resolve the cosmological ``Big Bang" singularity of Standard Cosmology and have hence attracted a lot of attention in the literature. Non-singular bounces were proposed a long time ago \cite{Tolman:1931zz}. They were studied in models motivated by approaches to quantum gravity such as modified gravity models \cite{Brustein:1997cv, Cartier:1999vk}, higher derivative gravity actions (see e.g. \cite{Mukhanov:1991zn, Tsujikawa:2002qc, Biswas:2005qr}), non-relativistic gravitational actions \cite{Brandenberger:2009yt, Cai:2009in, Gao:2009wn}, brane world scenarios \cite{Kehagias:1999vr, Shtanov:2002mb}, torsion gravity\cite{Cai:2011tc}, ``Pre-Big-Bang" \cite{Gasperini:1992em} and Ekpyrotic \cite{Khoury:2001wf} cosmology (in which case the conjectured bounces are classically singular), or loop quantum cosmology \cite{Bojowald:2001xe}. Bouncing cosmologies can also be obtained using arguments from super-string theory. For example, the String Gas Cosmology scenario \cite{Brandenberger:1988aj, Brandenberger:2008nx} may be embedded in a bouncing universe as realized in \cite{Biswas:2006bs}. Non-singular bounces may also be studied using effective field theory techniques by introducing matter fields violating certain energy conditions, for example non-conventional fluids \cite{Bozza:2005wn, Peter:2002cn}, quintom matter \cite{Cai:2007qw, Cai:2008qb, Cai:2009zp}, or ghost condensates \cite{Buchbinder:2007ad, Creminelli:2007aq, Lin:2010pf}. A non-singular bounce may also be achieved in a universe with open spatial curvature term (see e.g. \cite{Martin:2003sf, Solomons:2001ef}). A specific realization of a quintom bounce occurs in the Lee-Wick cosmology studied in \cite{Cai:2008qw}. Various original bounce models were reviewed in Ref. \cite{Novello:2008ra}.

In the context of studies of bouncing cosmologies it has been realized that fluctuations which are generated as quantum vacuum perturbations and exit the Hubble radius during a matter-dominated contracting phase lead to a scale-invariant spectrum of cosmological fluctuations today \cite{Wands:1998yp, Finelli:2001sr} (see also \cite{Starobinsky:1979ty} for an earlier discussion). This yields an alternative to inflation for explaining the current observational data, which is dubbed as the {\it matter bounce} (see e.g. \cite{Brandenberger:2010dk} for a recent review). However, in this scenario there are still unclear issues. For example, since scalar and tensor modes grow at the same rate in a matter-dominated phase of contraction, the tensor-to-scalar ratio is typically too large to be consistent with current observational data. Also, the possible generation of entropy and particles during the non-singular bounce phase needs to be studied in detail. In order to solve the first question, the scenario of matter bounce curvaton was proposed in \cite{Cai:2011zx}. In this paper, we will focus on the question of particle production in matter bounce cosmology.

The issue of particle production during the bouncing phase is important for various reasons. First of all, in the context of cold symmetric initial conditions and vacuum initial conditions for the fluctuations in a matter-dominated phase of contraction it is important to understand how the currently observed entropy of the universe can arise. In this paper we will explore the possibility that this radiative entropy is generated during the bounce phase, in analogy to how in inflationary cosmology the current entropy is generated during the ``reheating" phase. Secondly, it is important to show that the entropy produced in the initial stages of the bouncing phase (i.e. before the bounce point is reached) is not too large to disrupt the bounce. In many models of a non-singular bounce (e.g. the two matter field quintom bounce or the Lee-Wick bounce) the bounce is unstable to the presence of a substantial amount of radiation \cite{Karouby}, since radiation red-shifts as $a^{-4}$ whereas the regular and ghost matter fields whose interaction leads to the cosmological bounce red-shifts at most as $a^{-3}$, where $a(t)$ is the cosmological scale factor.

We will work in the context of realizations of the bounce using two matter scalar fields, one with regular kinetic term, the other with phantom sign \footnote{For readers worried about the ghost instability of this model \cite{Cline} we point out that we are treating our model as an effective field theory. There are realizations of the matter bounce which are, at least at the perturbative level, free from this ghost instability. This is the case e.g. for the cosmological bounce obtained in Ho\v{r}ava-Lifshitz gravity \cite{Brandenberger:2009yt} and ghost condensate bounce scenarios \cite{Buchbinder:2007ad, Creminelli:2007aq, Lin:2010pf}.}. A crucial ingredient of the bounce model which we will make use of here is the modeling of regular matter by an oscillating scalar field condensate. The oscillating scalar field condensate resembles the inflaton condensate during the preheating period during which the inflaton is coherently oscillating about the ground state of its potential energy. We know from studies of reheating in inflationary cosmology what a very interesting preheating dynamics describes the matter transfer between the inflaton and regular matter. In a similar way, we expect that the oscillations of the scalar field matter condensate in the matter bounce scenario can excite fluctuations and thus produce radiation. This is the topic we study in this paper.

Recently, the theory of preheating of the universe has become one of the most important issues in inflationary cosmology. During the preheating stage the energy is quickly transferred from the primordial inflaton scalar to matter fields which couple to the inflaton. This process occurs after the slow-roll period of inflation has ended, but earlier thermal equilibrium is established. The energy transfer from the inflaton to matter was initially analyzed using first order perturbation theory and discussed in terms of the decay  inflaton quanta into Standard Model particles \cite{Abbott:1982hn, Dolgov:1982th, Albrecht:1982mp}. However, it was realized that this analysis misses important effects due to the coherence of the inflaton condensate \cite{Traschen:1990sw}. The generation of matter in the inflaton condensate was then studied in a semi-classical analysis involving non-perturbative effects in Ref. \cite{Traschen:1990sw}. It was found that a parametric resonance instability plays a crucial role. These parametric resonance effects after inflation were further studied in Refs. \cite{Kofman:1994rk, Shtanov:1994ce} and the particle production process was then analyzed in detail in \cite{Kofman:1997yn}. In recent years, this topic has been extensively studied in the literature in the framework of inflationary cosmology \cite{Boyanovsky:1994me, Baacke:1996se, Cormier:2001iw}, and we refer to Refs. \cite{Bassett:2005xm} and \cite{Allahverdi:2010xz} for comprehensive reviews.

One of the interesting predictions made by preheating is that the metric fluctuations could be amplified due to entropy fluctuations\cite{BaVi,Finelli:1998bu}, even these fluctuations are outside the hubble radius. In the context of inflationary cosmology, it has been realized that entropy fluctuations can lead to an additional source of curvature fluctuations, a source which may in fact dominate \cite{Mollerach:1989hu, Linde:1996gt, Lyth:2001nq, Moroi:2001ct, Enqvist:2001zp}. This is now know as the ``curvaton mechanism" for generating fluctuations. Its application in matter bounce cosmology has been studied in Ref. \cite{Cai:2011zx}.

In analogy with inflationary preheating, we expect the oscillations of the matter condensate to generate fluctuations in fields which matter condensate couples to. This is the process we study in this paper. As we show, this process can indeed lead to the generation of radiation during the bounce. There are various constraints on this scenario. Firstly, the coupling of the matter condensate to the radiation field must be strong enough to allow for the preheating instability. A more stringent lower bound on the coupling arises if we demand that the preheating mechanism here has the strength to generate all of the radiation observed today.
On the other hand, there are also upper bounds on the strength of the coupling. Not too much radiation is allowed to be generated prior to the bounce point - otherwise the radiation would destabilize the bounce. Furthermore, the back-reaction of the produced particles is not allowed to be too strong to shut off the resonance via back-reaction. As we show here, there is a range of parameters where all of the above-mentioned constraints
can be satisfied.

The paper is organized as follows. In Section II we provide the basic equations of matter bounce cosmology. In Section III we study the preheating process of the simplest example in the framework of matter bounce cosmology. Explicitly, we study stochastic resonance with  weak and strong coupling coefficients, respectively, and find the resonance is generically more efficient than that in inflationary cosmology. Section IV is devoted to a discussion of the constraints on the model, and in Section V we show that the simplest version of preheating by letting the entropy field couple to the background scalar directly would spoil the scale-invariance of the primordial power spectrum in the matter bounce curvaton scenario. Instead, we can introduce a non-minimal coupling between the entropy field and the Ricci scalar so that the curvaton field can be thermalized as well due to gravitational interactions. Section VI presents a summary and discussion.

We make use of the convention $m_{pl} = 1/\sqrt{G}$ in the present paper.

\section{Basic picture of matter bounce cosmology}

We will discuss matter bounce scenarios in which the background matter which is responsible for the matter-dominated phase of contraction is a scalar field $\phi$ with standard kinetic term. In order to obtain a non-singular bounce in the context of an effective field theory in which space and time are described by the Einstein action, we need to introduce matter which violates the Null Energy Condition (NEC). There are various ways of achieving this. The details will not be important for our analysis, and therefore we simply use a general Lagrangian ${\cal L}_g$ as the Lagrangian of the matter which violates the NEC. We assume that in the initial phases of contraction the fields involved in the NEC violating sector make a negligible contribution to the total energy-momentum tensor. In order to get a bouncing background cosmology, the fields of ${\cal L}_g$ must become more important as the contraction of space proceeds, and eventually become equally important as the standard
matter at the bounce point.

As is done in analyses of inflationary reheating, we will model Standard Model matter as another scalar field $\chi$ minimally coupled to Einstein gravity $R$ and coupled to $\phi$ via an interaction potential $V(\phi, \chi)$. Thus, our action is
\begin{eqnarray}\label{action}
 S \, = \,  \int d^4x\sqrt{-g} \bigg[ \frac{R}{16\pi G}  &-& \frac{1}{2}\partial_\mu\phi\partial\phi
 - \frac{1}{2}\partial_\mu\chi\partial\chi \nonumber\\
 &-& V(\phi,\chi) + {\cal L}_g]~.
\end{eqnarray}
As a simple example, we take the potential to be
\begin{eqnarray}\label{potential}
 V(\phi,\chi) \, = \, \frac{1}{2}m^2\phi^2+\frac{1}{2}g^2\phi^2\chi^2~.
\end{eqnarray}

Our model is constructed such that there exists a nonsingular bounce. The bounce point occurs at the time $t_B$ which without loss of generality can be taken to be $t_B = 0$. We choose initial conditions such that long before the bounce point, the universe is contracting and dominated by a homogeneous condensate of the scalar field $\phi$. Thus, the equation of state averages to zero over time. During this period, the Hubble radius is decreasing linearly and $\dot{H} < 0$.

The model is constructed such that the contribution of ${\cal L}_g$ becomes more and more important as the contraction proceeds. The matter component making up ${\cal L}_g$ could be the Lee-Wick partner of $\phi$ \cite{Cai:2008qw}, it could be a condensate of ghost fields \cite{Lin:2010pf}, or it could be terms in an effective action involving higher order spatial derivatives \cite{Brandenberger:2009yt}. Independent of the specific mechanism, at a certain point in the contraction phase there the Hubble radius $|H|^{-1}$ will cease to decrease any further and a brief \footnote{The time scale of this bouncing phase (during which the equation of state of matter crosses the phantom divide twice, first decreasing below $w = -1$ and later - after the bounce point - rebounding to $w > -1$) is typically set by a high energy scale which enters into the construction of
${\cal L}_g$.} bouncing phase begins. During this period $\dot{H} > 0$. After the bouncing phase there comes a time when ${\dot H}$ vanishes, and after that the universe enters the expanding phase with a normal thermal history.

Returning to the question of initial conditions, we can take the contracting phase to be the mirror inverse of the expanding phase of Standard Cosmology, i.e. containing both matter and radiation, with radiation dominating over matter for $t > - t_{eq}$, where $t_{eq}$ is the usual time of equal matter and radiation for Standard Cosmology. If we take fluctuations to begin early in the contracting phase in their vacuum state, then these initial conditions will lead to a spectrum of curvature fluctuations at late times which is scale-invariant on large scales but makes a transition to a blue spectrum at scale which exit the Hubble radius after $t = - t_{eq}$. The observational constraints on this scenario were discussed in \cite{LiHong}. Alternatively, we can assume that there is no radiation at all in the contracting phase. In this case, the challenge is to produce the radiation which we currently observe by processes close to the bounce point. This is the scenario we study here.

We consider the metric of a homogeneous flat Friedman-Robertson-Walker space-time,
\begin{eqnarray}
 ds^2  \, = \, -dt^2 + a^2(t)d\vec{x}^2~,
\end{eqnarray}
where $t$ is the physical time, ${\vec{x}}$ denote the co-moving spatial coordinates and $a(t)$ is the cosmological scale factor,  and calculate the background evolution. The expansion rate of the universe, i.e., the Hubble parameter, is defined by $H\equiv\dot{a}/a$ and it and its time derivative are determined by the averaged Friedmann equations,
\begin{eqnarray}
 H^2 \, &=& \, \frac{8\pi G}{3}\langle\rho\rangle~,\\
 \dot{H} \, &=& \, -4\pi G\langle\rho+p\rangle~,
\end{eqnarray}
where $\rho$ and $p$ are the total energy density and pressure of the universe respectively.

\begin{figure}[htbp]
\includegraphics[scale=0.8]{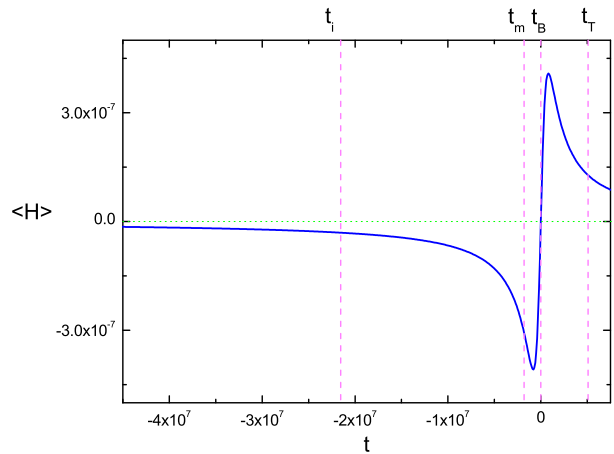}
\caption{Evolution of the Hubble parameter $H$ as a function of cosmic time $t$ in a matter bounce cosmology. In the numerical computation, the parameters $m$ and $g$ in the potential were chosen to be $m=10^{-6}m_{pl}$ and $g=10^{-4}$. The ghost part ${\cal L}_g$ part taken to as that of a Lee-Wick scalar \cite{Cai:2008qw} with mass scale to be $M=50m$. The times $t_i$, $t_m$, $t_B$, and $t_T$ denote the moment when stochastic resonance begins, the time when ${\dot H} = 0$, the bounce point, and the time of thermalization, respectively. The plot is in units of $m_{pl}$. } \label{Fig-st}
\end{figure}

In Fig. \ref{Fig-st} we numerically plot the evolution of the Hubble parameter in a nonsingular matter bounce cosmology (we use the specific model of \cite{Cai:2008qw} in which the Lee-Wick partner of $\phi$ yields the ghost Lagrangian ${\cal L}_g$. One can read off that the Hubble parameter is initially negative which implies a contracting universe. In this phase
\begin{eqnarray}\label{hubble}
 \langle H(t) \rangle \, = \, \frac{2}{3t} \, ,
\end{eqnarray}
where the pointed brackets indicate averaging. The background scalar condensate $\phi$ can be asymptotically expressed as an oscillating function
\begin{eqnarray}\label{phi}
 \phi(t) \, \simeq \, \tilde\phi(t)\sin mt~,
\end{eqnarray}
with a time dependent amplitude
\begin{eqnarray}\label{tphi}
 \tilde\phi(t) \, = \, \frac{m_{pl}}{\sqrt{3\pi}m|t|}~,
\end{eqnarray}

A very important lesson of this scenario is that the metric perturbations which originate as as quantum vacuum fluctuations on sub-Hubble scales are able to cross the Hubble radius during the contracting phase and evolve on super-Hubble scale into classical curvature fluctuations with a scale-invariant power spectrum \cite{Finelli:2001sr}. As was pointed out in \cite{Cai:2011zx}, isocurvature modes corresponding to massless scalar fields also obtain a scale-invariant power spectrum. Thus, with suitable choices on the parameters, the matter bounce model can fit the current CMB observations and explain the formation of large scale structure of our universe \cite{Cai:2008qw, Cai:2009hc}. In addition, a matter bounce cosmology predicts a particular type of primordial non-gaussianities with a sizable amplitude \cite{Cai:2009fn}. However, without quantum excitations of light matter fields, the universe described by a matter bounce would never arrive at the phase of thermal equilibrium, and thus there would be no  radiation phase and no process of entropy production in this model. Note that the same conclusion would hold in inflationary cosmology if there were no couplings between the inflaton and matter fields.

The analogy with inflation also tells us how to address this potential problem. We expect that the background field $\phi$ couples to other matter fields, such as the $\chi$ field introduced in Eq. (\ref{action}). As is well known, by virtue of the coupling in Eq. (\ref{potential}), one obtains a resonant production of $\chi$ particles for inflationary cosmology. One should expect a similar parametric resonance instability to occur in the matter bounce scenario. It is to a discussion of this issue to which we now turn.

\section{Stochastic Resonance of an Entropy Field in Bounce Cosmology}

In inflationary cosmology, the theory of preheating process has been intensively studied in the literature. This process occurs ubiquitously as a period of stochastic resonance in inflation models  when the inflaton starts to oscillate about its vacuum following the phase of slow-roll. As we motivated above, an analogous process is expected to occur near the bounce phase when the universe evolves from contraction to expansion. Thus, in the following we study the stochastic resonance of an entropy field in bounce cosmology due to the coherent dynamics of the $\phi$ condensate.

\subsection{The Setup}

To start, we study the quantum dynamics of the entropy scalar $\chi$ induced by the coherent oscillations of the condensate $\phi$, first  in the contracting phase, and then after the expanding period. Varying the action (\ref{action}) with respect to $\chi$, one gets the equation of motion of the inhomogeneous entropy field
\begin{eqnarray}
 \ddot\chi + 3H\dot\chi - \frac{\nabla^2}{a^2}\chi + g^2\phi^2\chi \, = \, 0~,
\end{eqnarray}
where we ignored the presence of metric perturbations. The effective mass of $\chi$ is induced by its interaction with $\phi$ as given in the interaction potential (\ref{potential}). Note that, in our simplest case, the bare  mass of the entropy field $\chi$ is taken to be zero. This is a good approximation provided the mass of $\chi$ is much less than that of the background field $\phi$.

Since the classical background solution for $\chi$ evolves proportional to $a^{-3/2}$ away from the bouncing phase (as shown e.g. in \cite{Cai:2011zx}), the investigation of parametric resonance can be simplified by re-scaling the entropy field $X=a^{3/2}\chi$. Following the formulae developed in Ref. \cite{Kofman:1997yn}, we can derive the following simpler form of the equation of motion for the Fourier modes of the entropy field $X_k$
\begin{eqnarray}\label{eomX}
 \ddot{X}_k + w_k^2X_k \, = \, 0~,
\end{eqnarray}
where the frequency $w_k^2$ is given by,
\begin{eqnarray}\label{wk2}
 w_k^2 \, = \, \frac{k^2}{a^2}+g^2\phi^2-\frac{9}{4}H^2-\frac{3}{2}\dot{H}~.
\end{eqnarray}
The last two terms of (\ref{wk2}) comes from the gravitational effects which cancel each other in a matter-dominated phase.

In a general case, the homogeneous solution to Eq. (\ref{eomX}) is composed of a negative frequency mode and a positive frequency mode as follows,
\begin{eqnarray}\label{eomXk}
 X_k(t) \, = \, \frac{1}{\sqrt{2w_k}} \bigg[ \alpha_k(t)e^{-i\int^tw_kdt} + \beta_k(t)e^{i\int^tw_kdt} \bigg]~,
\end{eqnarray}
and from (\ref{eomX}) the evolutions of these two coefficients $\alpha_k$ and $\beta_k$ satisfy the equations,
\begin{eqnarray}\label{alphabeta}
 \dot\alpha_k &=& \frac{\dot{w}_k}{2w_k}e^{2i\int^tw_kdt}\beta_k~, \nonumber\\
 \dot\beta_k &=& \frac{\dot{w}_k}{2w_k}e^{-2i\int^tw_kdt}\alpha_k~.
\end{eqnarray}

{F}ollowing the standard approach of canonical quantization, one obtains the conjugate momentum $\Pi_k$ for the field variable $X_k$:
\begin{eqnarray}
 \Pi_k \, &\equiv& \, \frac{\delta S}{\delta X_k} \nonumber\\
 &=& \, i\sqrt{\frac{w_k}{2}}\bigg[-\alpha_ke^{-i\int^tw_kdt}+\beta_ke^{i\int^tw_kdt}\bigg]~,
\end{eqnarray}
where we have used the equations (\ref{alphabeta}). Imposing the canonical normalized commutation relations
\begin{eqnarray}
 [X_k,\Pi_{k'}] \, = \, i\delta(w_k-w_{k'})~,
\end{eqnarray}
one finds that the coefficients $\alpha_k$ and $\beta_k$ coincide with the coefficients of the Bogoliubov transformation of the creation and annihilation operators in curved space-time. Thus, we impose the normalized initial conditions,
\begin{eqnarray}
\alpha_k^i \, = \, 1~,~~\beta_k^i \, = \, 0~,
\end{eqnarray}
at the beginning of the preheating period $t_i$. Moreover, the occupation number density for $\chi$ particles with co-moving wave number $k$  is given by
\begin{eqnarray}
 n_k \, = \, \frac{w_k}{2}\bigg(\frac{|\Pi_k|^2}{w_k^2}+|X_k|^2\bigg)-\frac{1}{2}~,
\end{eqnarray}
where the last term $\frac{1}{2}$ on the right hand side is a c-number used to cancel the zero-point energy.

There is a parametric resonance instability in the above cosmological system which can be described in terms of successive periodic scatterings of the $\chi_k$ modes in a parabolic potential, as first studied in \cite{Kofman:1997yn}. The resonance is effective if the condition
\begin{eqnarray} \label{rescond}
 |gm_{pl}| \, > \, \frac{3 \pi}{2} m
\end{eqnarray}
is satisfied (which comes from demanding that $q > 1$ is realized at some point during the oscillatory phase of $\phi$ - see the discussion below). In this case the total number density of the $\chi$ particles created during stochastic resonance can be estimated to be
\begin{eqnarray} \label{nchi}
 n_{\chi}(t) \, &=& \, \int(\frac{dk}{2\pi a})^3n_k(t) \nonumber\\
 &\simeq& \, \frac{k_i^3e^{2m\eta(t-t_i)}}{64\pi^2a^3\sqrt{\pi m \eta(t-t_i) + 2\pi^2}}~,
\end{eqnarray}
where $t_i$ denotes the moment when stochastic resonance begins (this time is shown in Fig. \ref{Fig-st}).

The value of $t_i$ in bounce cosmology is different from that in inflation. In inflationary cosmology, the preheating phase starts which $\phi$ takes on the largest value it has during its phase of oscillation, whereas in the matter bounce the preheating phase starts at the lowest value of $\phi$ when the efficiency condition for resonance is satisfied (this is the condition $q > 1$ to be discussed below). Explicitly, $t_i=\frac{\pi}{2m}$ for inflation while
\begin{eqnarray}
t_i \, = \, -\frac{gm_{pl}}{\sqrt{3\pi}m^2}
\end{eqnarray}
in bounce cosmology as will be explained in the following section.

Resonance takes place for all values of $k$ in certain resonance band. In the case of stochastic resonance (which is realized here), the dominant resonance band is the long wavelength band whose half width $k_i$ at the initial time $t_i$ is given by
\begin{eqnarray}
 \frac{k}{a}_i \, \simeq \, \sqrt{gm|\tilde\phi_i|} \, .
\end{eqnarray}
Its physical wave number approximately equals to $\sqrt{gmm_{pl}/5}$ in inflationary cosmology but $m$ in bounce cosmology. The parameter $\eta$ is the largest value of the Floquet (growth) index within the instability band, and its effective value is
\begin{eqnarray}
 \eta \, \simeq \, 0.18 \, \sim \, 0.1 \, .
\end{eqnarray}

\subsection{Broad Resonance with Weak Coupling}

Recall that the cosmological scalar $\phi$ oscillates around its vacuum and the universe evolves as matter dominated one in the contracting phase. Its background evolution is described by Eqs. (\ref{hubble}), ({\ref{phi}}) and the amplitude is given by (\ref{tphi}). It is useful to use, instead of cosmic time $t$, the number of oscillations of the homogeneous scalar $\phi$, which is (beginning the count at the time $t_i$ when resonance first becomes effective)
\begin{eqnarray}
 N(t) \, = \, \frac{m(t-t_i)}{2\pi}~.
\end{eqnarray}

In the analytic computation of a process of stochastic resonance, it is convenient to define a resonance parameter
\begin{eqnarray}
 q \, \equiv \, \frac{g^2\tilde\phi^2}{m^2}~.
\end{eqnarray}
Studies of reheating in inflationary cosmology \cite{Kofman:1997yn, Allahverdi:2010xz} have taught us that in a dynamical space-time only the broad resonance regime with $q>1$ can lead to a significant enhancement of the generation of the $\chi$ particles and of the corresponding fluctuations on super-Hubble scales. Since in the framework of bounce cosmology the universe starts its evolution from a very low energy state in a contracting phase, the value of $q$ is initially much smaller than $1$ which indicates broad resonance cannot take place in
most of this regime. However, along with the contraction, the amplitude of the background scalar $\phi$ increases  gradually, and consequently, $q$ will approach to the boundary $q=1$ and then cross over. That moment corresponds to the starting time $t_i$ of broad resonance in bounce cosmology, which -  if we follow the solution obtained in Eq. (\ref{tphi} - is given by
\begin{eqnarray}
 t_i \, = \, -\frac{gm_{pl}}{\sqrt{3\pi}m^2} \, ,
\end{eqnarray}
with $\tilde\phi_i=\frac{m}{g}$).

As long as we can neglect the back-reaction of the $\chi$ particles produced during the resonance on the background $\phi$ condensate, the condensate will oscillate with increasing amplitude $\tilde\phi$ until it reaches the value
\begin{eqnarray}
 \tilde\phi \, = \, \frac{2m_{pl}}{\sqrt{3\pi^3}} \, ,
\end{eqnarray}
which occurs at the time
\begin{eqnarray}
 t_m \, = \, -\frac{\pi}{2m}
\end{eqnarray}
in the contracting phase, and then the universe will enter the bouncing phase with the broad resonance being replaced by a process of gravitational resonance as studied in Sec. V. After the bounce, there is one another period of broad resonance in the expanding phase, which is a reversal of the process in the contracting phase.

Therefore, during the process of broad resonance with weak coupling in bounce cosmology, the scalar $\phi$ experiences the following number of oscillations
\begin{eqnarray}
 N \, \simeq \, \frac{gm_{pl}}{3\pi m}~,
\end{eqnarray}
which is double of that obtained in usual inflationary cosmology. For example, for the set of parameters $m = 10^{-6}{m_{pl}}$ and $g = 10^{-4}$, the above estimate gives $N\simeq10$. Correspondingly, the increase in the occupation number density for the $k=0.1 m$ mode of $\chi$ is given by
\begin{eqnarray}
 \Delta \ln n_k \, \simeq \, 4\pi\eta_k N-\ln2 \, \simeq \, 15
\end{eqnarray}
if we use the value $\eta_k \simeq 0.13$.

\begin{figure}[htbp]
\includegraphics[scale=0.8]{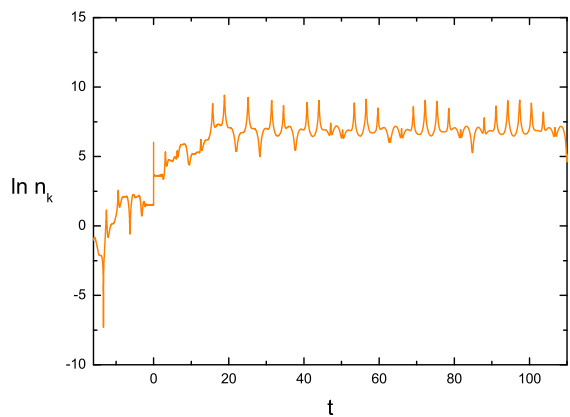}
\caption{Evolution of the occupation number density $n_k$ during the process of broad resonance with weak coupling as a function of cosmic time (horizontal axis). The time axis is displayed in units of ${m}^{-1}$. The parameters $m$ and $g$ were chosen to be $m = 10^{-6} {m_{pl}}$ and $g = 10^{-4}$. In the computation, the mode was chosen as $k=10^{-7} {m_{pl}}$ with resulting initial conditions $X_k= 0.7386$ and $\dot{X}_k = 0$.} \label{fig:nk}
\end{figure}

\begin{figure}[htbp]
\includegraphics[scale=0.8]{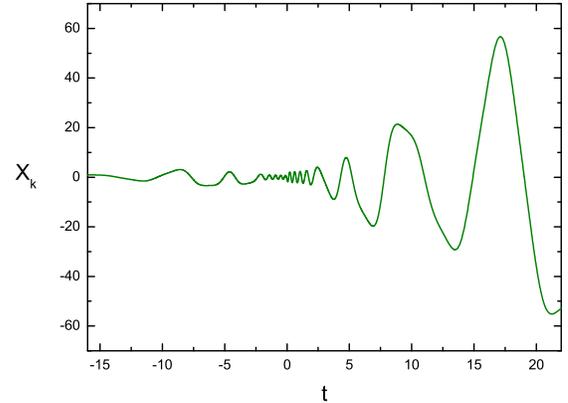}
\caption{Evolution of the variable $X_k$ during the process of broad resonance with weak coupling. The horizontal axis is cosmic time. The initial values of the background parameters are the same as in Figure \ref{fig:nk}.} \label{fig:Xk}
\end{figure}

To get a better understanding of this process, we solve the perturbation equation (\ref{eomXk}) numerically and show the results in Figs. \ref{fig:nk} and \ref{fig:Xk}. We find that the analytic estimate is in good agreement with the numerical results shown in these figures. One can conclude that the main mechanism of  broad resonance with weak coupling is similar to that in inflationary cosmology, but that the oscillation number is doubled since there are now two time periods of resonance as a consequence of the background evolution.

\subsection{Pre-Bounce Resonance with Strong Coupling}

As we have analyzed in the above subsection, when $g \gtrsim \sqrt{2 \pi} m/m_{pl}$, our model will experience a period of broad parametric resonance as occurs in inflationary preheating. Note that we have so far neglected the back-reaction due to the presence of $\chi$ particles. Once the value of  the coupling constant $g$ is sufficiently large, this back-reaction can no longer be neglected. This back-reaction can change the structure of the resonance, and it can also change the background evolution. We will study the first issue during the pre-bounce resonance in the following.

The rough criterion for the back-reaction to have a negligible effect on the structure of the resonance is to demand that the change $\Delta{m}_\phi^2$ in the mass of $\phi$ as a consequence of interactions with the produced $\chi$ particles be smaller than its intrinsic mass square $m^2$. In the Hartree approximation, the change in the mass of the background scalar $\phi$ due to the back-reaction of $\chi$ quanta is expressed as,
\begin{eqnarray}
 \Delta{m}_\phi^2 \, = \, g^2 \langle\chi^2\rangle~.
\end{eqnarray}
Here, $\langle\chi^2\rangle$ indicates the  quantum expectation value of the square of the entropy field, which in turn is given by
\begin{eqnarray} \label{chiexp}
 \langle\chi^2\rangle \, = \, \int_0^\infty dk \frac{k^2}{2\pi^2a^3}|X_k|^2~.
\end{eqnarray}

The preheating process in the contracting phase can be divided into two periods. In the first the level of excitation of $\chi$ particles is low enough to be able to neglect the back-reaction effects, and the frequency of oscillations of the scalar $\phi$ is still determined by the mass $m$. Recall  that we used the occupation number density (\ref{nchi}) to describe the production of $\chi$. As can easily be seen, the expectation value (\ref{chiexp}) is related to the number density of $\chi$ particles via
\begin{eqnarray}
 \langle\chi^2\rangle \, \sim \, \frac{a}{k_i} n_{\chi} \, ,
\end{eqnarray}
where $k_i$ is the half width of the dominant resonance band and which before the bounce corresponds to a physical wave number $(k/a)_i\simeq m$. As the generation of $\chi$ particles, the back-reaction effect starts to be important, leading to an end of the first stage of resonance at a moment $t_{p1}$ when
\begin{eqnarray}
 n_{\chi}(t_{p1}) \, = \, \frac{m^2}{g} \tilde\phi(t_{p1})~,
\end{eqnarray}
which yields
\begin{eqnarray}
 t_{p1} \, \simeq \, -\frac{{\cal W}(y)}{2\eta{m}}~,~~
 y \, = \, \frac{g^{\frac{5}{2}}\eta^{\frac{1}{2}}m_{pl}^{\frac{1}{2}}}{32\pi^3m^{\frac{1}{2}}} e^{\frac{2g\eta{m}_{pl}}{3m}}~,
\end{eqnarray}
where ${\cal W}$ is a Lambert W-function.

Consequently, the oscillation number of $\phi$ in the first stage of preheating can be estimated by the following approximate relation:
\begin{eqnarray}
 N_{p1} \, \simeq \, \frac{gm_{pl}}{6\pi{m}} - \frac{{\cal W}(y)}{4\pi\eta}~.
\end{eqnarray}
In the first stage of pre-bounce resonance, the amplitude of the scalar $\tilde\phi$ increases and thus leads to $q\gg1$.

The preheating process will enter a second stage if $t_{p1}<t_m$, i.e. if back-reaction becomes important before the amplitude of $\phi$ has reached its maximal value. This gives a criterion for the coupling constant $g$: a second stage of preheating will occur if $g$ is larger than the following critical value:
\begin{eqnarray}
 g_c \, \simeq \, \frac{15m}{4\eta{m}_{pl}}{\cal W}(\frac{4\eta^{\frac{4}{5}}m_{pl}^{\frac{4}{5}}}{m^{\frac{4}{5}}}e^{\frac{2}{5}\pi\eta})~.
\end{eqnarray}

There is one further condition on $g$: if $g$ is too large, then back-reaction cannot be neglected at the initial time when the resonance condition becomes satisfied. This will occur if $t_{p1} > t_i$ (recall that both of these times are negative). We will call the related critical value of $g$ by $g_i$.  Thus, we require $g\lesssim g_i$ with
\begin{eqnarray}
 g_i  \, \simeq \, 4\pi^{\frac{6}{5}} \bigg(\frac{m}{\eta m_{pl}}\bigg)^{\frac{1}{5}}~.
\end{eqnarray}

\begin{figure}[htbp]
\includegraphics[scale=0.8]{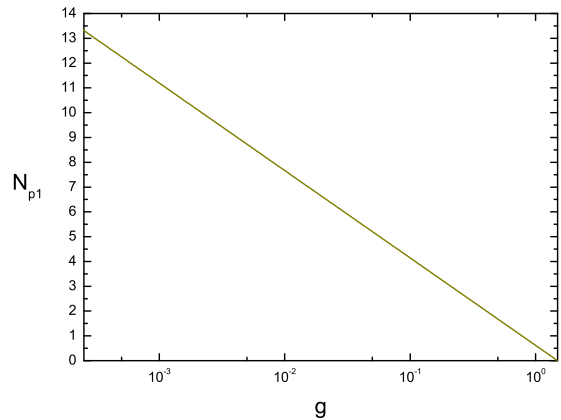}
\caption{The number of oscillations $N_{p1}$ in the first stage of pre-bounce resonance as a function of the coupling constant $g$. In the calculation we took $m=10^{-6}m_{pl}$ and $\eta=0.13$.}
\label{fig:Np1}
\end{figure}

If we choose $m=10^{-6}m_{pl}$ and $\eta=0.13$, we find that in order to have successful preheating beginning in a first stage during which the $\chi$ back-reaction is negligible and leading on to a second stage when the $\chi$ particles dominate the mass of $\phi$,  we require the coupling parameter $g$ to be bigger than $g_c \simeq 2.5\times10^{-4}$ but less than $g_i\simeq1.5$.

We have performed a numerical estimate of the oscillation number $N$ as a function of $g$ and the results are displayed in Fig. \ref{fig:Np1}. One may notice that there are no oscillations of the scalar $\phi$ when $g$ is larger than of order $O(1)$. This is understandable since $g > 1$ corresponds to a very strongly coupled system and so the present perturbative treatment is no longer valid. We will not consider such large values of $g$ any further in this work.

{F}inally, let us estimate the amount of preheating in the second stage, during which the frequency of the $\phi$ oscillations is no longer determined by $m$ but instead by $g\chi_{\rm rms}$. Therefore, the resonance parameter is effectively modified to be
\begin{eqnarray}
 q_{\rm eff} \, = \, \tilde\phi^2/\chi_{\rm rms}^2 \, .
\end{eqnarray}
The stage terminates at the moment $t_{p2}$ when the effective resonance parameter decreases to $1$. As a consequence, similar to what is done in the analysis in inflationary preheating, we obtain the following oscillation number in the second stage of pre-bounce resonance:
\begin{eqnarray}
 N_{p2} \, \simeq \, \frac{\ln(2\sqrt{2}q_{r1}^{\frac{1}{4}})}{4\pi\eta}~,
\end{eqnarray}
and correspondingly the number density of $\chi$ particles is given by
\begin{eqnarray}
 {n_{\chi}}_{p2} \, \simeq \, {n_{\chi}}_{p1} e^{4\pi\eta{N}_{p2}} \, .
\end{eqnarray}
Inserting the values of parameters chosen in Fig. \ref{fig:Np1}, one finds that $N_{p2} \lesssim 1$ in most of the allowed regime for $g$. This result indicates that once the back-reaction of $\chi$ particles dominates over the background universe in the contracting phase, there is no longer a broad resonance band after the bounce.

To briefly summarize: in this section we have analyzed the parametric resonance of the entropy field $\chi$ arising in the simplest model in the framework of a nonsingular bounce cosmology. We find that there exist two different resonance regions (in time): one is a broad resonance region with weak coupling which starts in the contracting phase once the amplitude of the oscillating matter condensate field $\phi$ exceeds some critical value, and finally ends in the expanding phase once the amplitude falls again below this critical value. The total oscillation number during this broad resonance interval is twice that obtained in inflationary cosmology.
Around the bounce point there may be a second region of broad resonance with a relatively strong coupling. This second region arises if the back-reaction of the entropy particles on the background condensate becomes important. If this back-reaction becomes important, then the existence of the part of the first resonance interval after the bounce will be affected.

\section{Constraints on $g$}

There are two additional constraints on the model parameters which we have not yet discussed. First, the resonance has to be weak enough such that the $\chi$ particles (whose energy density grows as $a^{-4}$ and thus behave as radiation) do not dominate the energy density at the bounce time (bounce time computed without taking into account the back-reaction of the $\chi$ particles). If they were to dominate, the bouncing background would be destabilize and the universe would collapse to a Big Crunch singularity instead of bouncing. Secondly, if we want preheating to explain the origin of the current entropy, then we require that at the end of the resonance
period in the expanding phase the energy density in $\chi$ particles becomes dominant. In this section we discuss these and the other conditions on the coupling constant $g$.

First of all, we recall from (\ref{rescond}) that no broad resonance at weak coupling can take place unless
\begin{eqnarray}\label{gresonance}
 g  \, > \,  \frac{3\pi m}{2m_{pl}}~.
\end{eqnarray}
If we want to back-reaction of $\chi$ particles not to cut off the resonance before the bounce point, we have an upper bound on $g$
\begin{eqnarray}\label{gstrong}
 g  \, <  \, \frac{15m}{4\eta m_{pl}} {\cal W}(\frac{4\eta^{\frac{4}{5}}m_{pl}^{\frac{4}{5}}}{m^{\frac{4}{5}}}e^{\frac{2}{5}\pi\eta})~.
\end{eqnarray}

Next, we turn to the energy density conditions mentioned at the beginning of this section. The energy density of the entropy field $\chi$ can easily be obtained by integrating the number density times mode energy of $\chi$ modes over the dominant instability band. Making use of (\ref{nchi}) we obtain
\begin{eqnarray}\label{rho_chi}
 \rho_{\chi}(t) \, = \, \rho_{\chi}(t_i) (\frac{a_i}{a})^4 \frac{e^{2m\eta(t-t_i)}}{\sqrt{1+\frac{m\eta}{2\pi}(t-t_i)}}~,
\end{eqnarray}
where $a_i$ denotes the scale factor of the universe at the beginning moment of preheating. Note that, the second term of rhs of (\ref{rho_chi}) is arisen from the redshift effect, and the third term is due to the parametric resonance effect of preheating.

Now we compare the energy density of entropy field and that of background scalar at the moment $t_m$ when the background scalar ceases its oscillations in the contracting phase. We require $\rho_\chi < \rho_\phi$ at this moment, and thus obtain an upper bound on the coupling parameter, which takes the form
\begin{eqnarray}\label{gupper}
 g  \, < \,  \frac{13m}{4\eta m_{pl}}{\cal W}(\frac{3\pi\eta^{16/13}m_{pl}^{12/13}}{m^{12/13}})~.
\end{eqnarray}

Next, we calculate $\rho_\chi$ and $\rho_\phi$ at the end of the parametric resonance time interval (time $t_f$) in the expanding phase (neglecting the possible back-reaction of the $\chi$ particles on the background). If $\rho_\chi > \rho_\phi$ at $t_f$ then the universe will enter a radiation-dominated period after the bounce. As a consequence, we get a lower bound on the coupling parameter, which is
\begin{eqnarray}\label{glower}
 g  \, > \,  \frac{3m}{2\eta m_{pl}}{\cal W}(\frac{19\pi^{3/2}\eta m_{pl}}{3m})~.
\end{eqnarray}

\begin{figure}[htbp]
\includegraphics[scale=0.3]{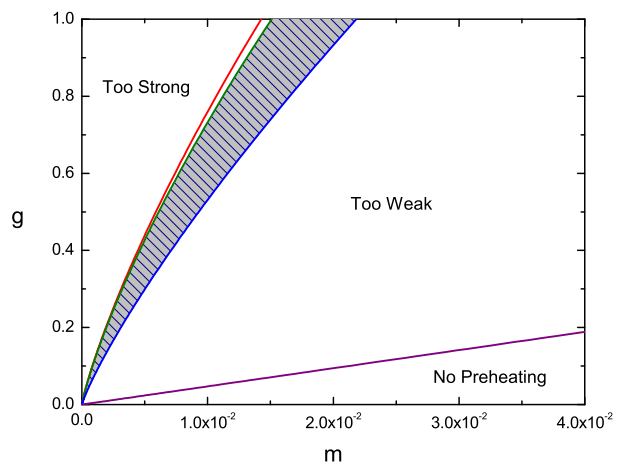}
\caption{The constraints on the parameter space of the coupling constant $g$. In the calculation we took $\eta=0.13$ and the mass of background field $m$ is in units of $m_{pl}$.}
\label{fig:constraint}
\end{figure}

Combining (\ref{gresonance}), (\ref{gstrong}), (\ref{gupper}) and (\ref{glower}), we find that the parameter space of the coupling constant $g$ is tightly constrained by these inequalities. Specifically, in Fig. \ref{fig:constraint} we plot the contours constraining the viable parameter space for $g$ In the region above the red line, $g$ is too strong so that the backreaction of $\chi$ dominates over the background before the bounce point. The requirement of controllable growth of $\chi$ particles before the bounce imposes an upper bound on $g$ which is shown by the green line as analyzed in (\ref{gupper}). The blue line corresponds to the lower bound on $g$ obtained by demanding that the universe be dominated by radiation after the preheating phase. Finally, the regime below the purple line does not allow the occurrence of parametric resonance.

\section{Geometric Preheating and the Scenario of Matter Bounce Curvaton}

The same two scalar field model as studied here was used in Ref. \cite{Cai:2011zx} to address one of the potential problems of the matter bounce scenario, namely the fact that a simple matter bounce typically produces a ratio of tensor to scalar metric fluctuations which is of order unity and already in marginal conflict with data. The idea is to introduce a light field $\chi$ to seed  isocurvature fluctuations. The infrared modes of the isocurvature fluctuations can obtain a kinetic amplification and are converted to curvature perturbations in the bouncing phase. This so-called matter bounce curvaton scenario requires that  the coefficient $g$ of the coupling between the curvaton field and the background scalar has to be very small to ensure that the isocurvature fluctuations obtain a nearly scale-invariant. In fact, the condition on $g$ derived in \cite{Cai:2011zx} is the opposite of the criterium (\ref{rescond}) derived here which is required to obtain broad resonance.

Fortunately, these two different pictures can be unified by introducing a non-minimal coupling for the curvaton field. From the point of view of quantum field theories in curved space-time, non-minimal coupling terms often arise to cancel the divergencies which are widely encountered in the ultraviolet limit of renormalization group flows of cosmological scalars. As a consequence, we expect an existence of a non-minimal coupling for the curvaton field in the matter bounce can provide an alternative channel for parametric resonance. The same idea applied to inflationary cosmology has been analyzed in \cite{Bassett:1997az, Tsujikawa:1999jh} and was called  ``geometric preheating". In the following we explore the possibility of geometric preheating in the matter bounce curvaton scenario.

We consider the curvaton field $\chi$ non-minimally coupling to the Ricci scalar $R$ through the following term,
\begin{eqnarray}
 V_R \, = \, \frac{1}{2}\mu^2 R\chi^2~.
\end{eqnarray}
In the matter-dominated contracting universe we have
\begin{eqnarray}
 R \, &=& \, 6\dot{H}+12H^2\nonumber\\
 &\simeq& \, 8\pi\frac{m^2}{m_{pl}^2}\phi^2,
\end{eqnarray}
where we have made use of time averaging  and ignored the back-reaction of the entropy field. As a consequence, the frequency of the entropy field in Fourier space is given by
\begin{eqnarray}\label{wk2g}
 w_k^2  \, \simeq \, \frac{k^2}{a^2} + g^2\phi^2+8\pi\mu^2\frac{m^2}{m_{pl}^2}\phi^2~.
\end{eqnarray}
One notices that, even the coefficient of direct coupling between the curvaton $\chi$ and the background scalar $\phi$ is negligible, parametric resonance is still able to be taken place if the effective resonance parameter is larger than unit at the moment of $t_i$ as studied in the previous section. The condition for geometric preheating to arise in the matter bounce curvaton scenario is
\begin{eqnarray}
 \mu \, \gtrsim \, \frac{1}{8\sqrt{2\pi}} \simeq 0.05~.
\end{eqnarray}
Comparing this with the criterion of geometric preheating in inflation (which is $\mu \gtrsim \frac{\sqrt{3}\pi}{4\sqrt{2}} \simeq 1$), we see that the requirement in our model is much relaxed.

Note that the above estimation is obtained based on the following approximation: we only considered the time-averaged effect of the background evolution, and ignored the back-reaction of both the entropy field and the metric fluctuations. Thus, our result only give a qualitative guideline. It is necessary to do a detailed quantitative analysis of this scenario. Moreover, a general model involving non-minimal coupling to background fields might be interesting \footnote{Non-minimal matter bounce cosmology was studied in \cite{Qiu:2010ch}.}.
We would like to leave these issues to followup work.

\section{Conclusions and Discussion}

In this paper we have studied the theory of preheating in the framework of a nonsingular bouncing cosmology with a matter-dominated contracting phase. Important in our analysis is that the contracting phase is dominated by an oscillating scalar mater field condensate. We have seen that preheating is driven by a period of stochastic parametric resonance and can in principle explain the origin of the post-bounce entropy. Our analysis does
not depend much on the precise mechanism which realizes the non-singular bounce since the preheating takes place before the terms yielding the non-singular bounce become very important.

The epoch of preheating takes place near the bouncing phase after the amplitude of the condensate has had time to grow to a sufficiently large value for the parametric resonance parameter $q$ to become larger than $q = 1$. Depending on the magnitude of the coupling coefficient between the condensate and the entropy field $\chi$, preheating will proceed either throughout the whole pre- and post-bounce intervals during which $q > 1$,  or it will end before the bounce. When the coupling is weak, the process of stochastic resonance is similar to that in inflation, but the oscillation number is doubled. Therefore, compared with inflationary cosmology, there will be more particles of the entropy field excited in a bouncing cosmology. In the case of strong coupling, the back-reaction of the entropy field cannot be neglected and leads to an earlier cutoff for the time interval of
stochastic resonance. For a typical parameter values of our model, it is found that parametric resonance terminates before the universe arrives at the bounce point.

We also have studied an extended picture of preheating in which the entropy field couples to the background scalar field mainly gravitationally. Explicitly, the entropy field couples to the Ricci scalar non-minimally and thus can also give rise to parametric resonance. In this approach, we are able to accommodate the entropy generation studied here with the matter bounce curvaton scenario \cite{Cai:2011zx}. According to our qualitative estimates, the condition for the entropy field to preheat geometrically in a matter bounce scenario is slightly easier than in the case of inflation.

Note that we have not addressed the interesting question of how thermalization after preheating takes place in the context of a nonsingular bouncing universe. This thermalization process involves a lot of non-perturbative effects, and thus calls for corresponding non-perturbative analyses. Some numerical studies in the framework of inflationary cosmology were performed in recent years \cite{Khlebnikov:1996mc, Prokopec:1996rr, Felder:2000hq}, and it was found that the evolution of the entropy field enters a regime of turbulent scaling\cite{Micha:2002ey}. We expect that a similar phase could also occur in the matter bounce model. However, the most distinctive difference between preheating in inflation and matter bounce is the initial state. We expect that this could yield new results for thermalization of a bouncing universe.

To conclude, we would like to highlight the importance of our analysis. The study of preheating has a lot of applications to other topics, e.g. to topological defect production \cite{Khlebnikov:1998sz, Parry:1998de}, primordial magnetic fields \cite{Calzetta:2001cf, Boyanovsky:2002wa, DiazGil:2007dy}, induced non-Gaussianities
\cite{Enqvist:2004ey, Barnaby:2006cq, Chambers:2007se, Kohri:2009ac, Byrnes:2008zz}, preheating with non-standard kinetic terms \cite{Lachapelle:2008sy}, and so on. The mechanism of preheating in bounce cosmology, since the initial condition is modified when compared to inflationary preheating, could provide a new window to explore early universe phenomenology combined with particle physics. This process is rather robust and ought to be considered in all bounce models.

\begin{acknowledgments}

YC is supported by funds of department of physics at Arizona State University. The research of RB is supported in part by a NSERC Discovery Grant at McGill and by funds from the Canada Research Chairs program. The research of XZ is supported in part by the National Science Foundation of China under Grants No. 10533010 and 10675136, by the 973 program No. 2007CB815401, by NSFC No. 10821063, and by the Chinese Academy of Sciences under Grant No. KJCX3-SYW-N2. Two of us (YC and RB) wish to thank Professor Xinmin Zhang and the Theory Division of the Institute for High Energy Physics for hospitality during the period when the draft of this paper was finalized.
\end{acknowledgments}

\end{document}